\documentclass[10pt,letterpaper]{article}
\usepackage[top=0.85in,left=1.5in,footskip=0.75in,marginparwidth=2in]{geometry}

\usepackage[utf8]{inputenc}

\usepackage{cite}

\usepackage{nameref,hyperref}

\usepackage{amsmath,amssymb,amsfonts}
\usepackage{float}
\usepackage[lined,ruled,linesnumbered]{algorithm2e}

\usepackage{algorithmic}
\usepackage{graphicx}
\usepackage{textcomp}
\def\BibTeX{{\rm B\kern-.05em{\sc i\kern-.025em b}\kern-.08em
    T\kern-.1667em\lower.7ex\hbox{E}\kern-.125emX}}
\usepackage{booktabs}
\usepackage{array}
\usepackage{float}
\usepackage{multirow}

\def\BibTeX{{\rm B\kern-.05em{\sc i\kern-.025em b}\kern-.08em
    T\kern-.1667em\lower.7ex\hbox{E}\kern-.125emX}}

\usepackage{multirow}

\usepackage[labelformat=simple]{subcaption}

\DeclareCaptionLabelFormat{subcaptionlabel}{\normalfont(\textbf{#2}\normalfont)}
\usepackage{microtype}
\DisableLigatures[f]{encoding = *, family = * }

\raggedright
\usepackage{changepage}

\usepackage[aboveskip=1pt,labelfont=bf,labelsep=period,singlelinecheck=off]{caption}

\makeatletter
\renewcommand{\@biblabel}[1]{\quad#1.}
\makeatother

\usepackage{lastpage,fancyhdr,graphicx}
\usepackage{epstopdf}
\fancyheadoffset[L]{2.25in}
\fancyfootoffset[L]{2.25in}

\usepackage{color}

\definecolor{Gray}{gray}{.25}

\usepackage{graphicx}

\usepackage{sidecap}

\usepackage{wrapfig}
\usepackage[pscoord]{eso-pic}
\usepackage[fulladjust]{marginnote}
\reversemarginpar

\begin{document}
\vspace*{0.35in}

\begin{flushleft}
{\Large
\textbf\newline{Quality of Service (QoS) Modelling in Federated Cloud Computing}
}
\newline
\\
Kun Ma\textsuperscript{1,*},
Antoine B. Bagula\textsuperscript{1,2},
Olasupo O. Ajayi\textsuperscript{1}
\\
\bigskip
\bf{1} Department of Computer Science,University of the Western Cape, Bellville, South Africa
\\
\bf{2} ISAT Lab, Computer Science Department, University of the Western Cape, South Africa
\\
\bigskip
* makuning@126.com

\end{flushleft}

\section*{Abstract}
Building around the idea of a large scale server infrastructure with  a potentially large number of tailored resources, which are capable of interacting to facilitate the deployment, adaptation, and support of services,  cloud computing  needs to 
frequently reschedule and manage various application tasks in order to accommodate  the requests of a wide range and number of users. One of the challenges of cloud computing is to support and manage  Quality-of-Service (QoS) by designing efficient 
techniques for the allocation of  tasks between users and the cloud virtual resources, as well as assigning virtual resources to the cloud physical resources. The migration of virtual resources across physical resources is another challenge that requires considerable attention; especially in federated cloud computing environments wherein, providers might be willing to offer their unused resources as a service to the federation(cooperative allocation) and  pull back these resources for their own use when they are needed (competitive allocation). This paper revisits the issue of QoS in cloud computing  by formulating and presenting i) a multi-QoS  task allocation model for the assignment  of users' tasks to virtual machines and ii) a virtual machine migration model for a federated cloud computing environment by considering cases where resource providers are operating in cooperative or competitive mode.  A new differential evolution (DE) based binding policy for task allocation  and  a novel virtual machine model are proposed  as solutions for the problem of QoS support in federated cloud environments. The experimental results show that the proposed solutions improved the quality of service in the cloud computing environment and reveal the relative advantages of operating a mixed cooperation and competition model in a federated cloud environment.


\section*{Introduction}
Building around the idea of a large scale server infrastructure with  a potentially large number of tailored resources, cloud computing has developed as a new type of business model that promises  the supply of virtually unlimited capacities and services in a scalable, pay-as-you-go fashion. A cloud computing environment includes a  lot of virtualized dynamic computing resources capable of providing all kinds of computing and data storage services to users through the network. At the same time, the cloud system needs to be able to effectively coordinate distributed resources belonging to different agencies and individuals with the objective of  integrating and sharing large heterogeneous resources, as well as providing unified user access interfaces to external users. 
With the cloud, users simply connect to a network and have access to large and seemingly unlimited computing environments. The specifics of which resources are allocated to the user or the exact deployment location of these resources are abstracted from the user \cite{Laili}. 

Virtualisation is a technology that abstracts the physical infrastructure to provide virtualised resources called virtual machines (VMs) for advanced applications \cite{Zhang}. In cloud computing, virtualisation provides the ability to aggregate computing resources from a server cluster and assign VMs to jobs as needed. For each task execution request, a new VM can be initialised or the task can be assigned to an existing VM of the same user \cite{Rimal}. The maturity of the Cloud computing and continued adoptation of the technology have resulted in tremendous increase in the number and scale of cloud data centres. These however have a negative effect, as the amount of energy consumed have also grown astronomically and have now become a prominent issue that needs to be tackled\cite{Mell}. According to statistics, in 2011, data centre consumed about 1.4\% of the total global electricity consumption. This was also couped with a compound annual growth rate of 4.4\%, a figure which was much higher than the estimated global demand growth of 2.1\% \cite{Kapur}. As a result, energy consumption in cloud data centres is becoming a  challenging issue of great concern~\cite{Beloglazov}.

Beyond energy consumption, of direct impact on the users is Quality of service (QoS). In recent years, the importance of QoS in cloud computing is now more important than it previously was. There is now the need for Cloud systems to support per-user and per-application QoS requirements when providing services. These are requirements in which requesting users/applications include their QoS requirements when requesting services from the remote Cloud  resources \cite{Lin} and stiff penalities are attached to non-compliance.  The degree of QoS satisfaction is an important parameter that measures the performance of any scheduling strategy used in a cloud computing environment. Therefore, in cloud computing environments, the task scheduling policies usually consider the tasks' QoS target requirements  and the performance parameters of various available resources, in order to effectively match a large number of application tasks  to the available computing resources~\cite{Hirai} \cite{Pedersen}.

Service Level Agreements (SLA) between the cloud service providers (CSPs) and cloud users are used to define the QoS to be provided to users and the expectations of the users. Information such as throughput, reliability, blocking probability and response time, payment process and SLA violation penalties, etc. are often documented in the SLA. Research on how to dynamically schedule virtual machine resources to reduce overall system over-utilization while keeping SLA level is still a current hot spot in cloud computing research and practice. \cite{Buyya}

While cloud computing holds a lot of promise for enterprise computing, the single-cloud provider model raises several issues, some of them related to the limited scalability which arises when the cloud usage rate increases. Other issues are related to the absence of built-In Business Service Management Support leading to enterprises looking at transforming their IT operations to cloud-based technologies facing a non-incremental and potentially disruptive step. The lack of interoperability leading to the inability to scale through business partnerships across clouds providers and preventing small and medium cloud infrastructure providers from entering the cloud provisioning market is another issue. Federated cloud computing has the potential to address these issues as it is built around different cloud providers collaborating by: i) sharing their resources while having each of them remaining an independent autonomous entity by keeping thick walls between them by and ii) having the applications running in a cloud of clouds while being unaware of the location by having virtual local networks being designed and implemented to enable  the inter-application components to communicate and iii)  having cloud providers differentiate from each other in terms of cost 
and trust level. 

\begin{figure}[h]
\centering
\includegraphics[scale=0.5]{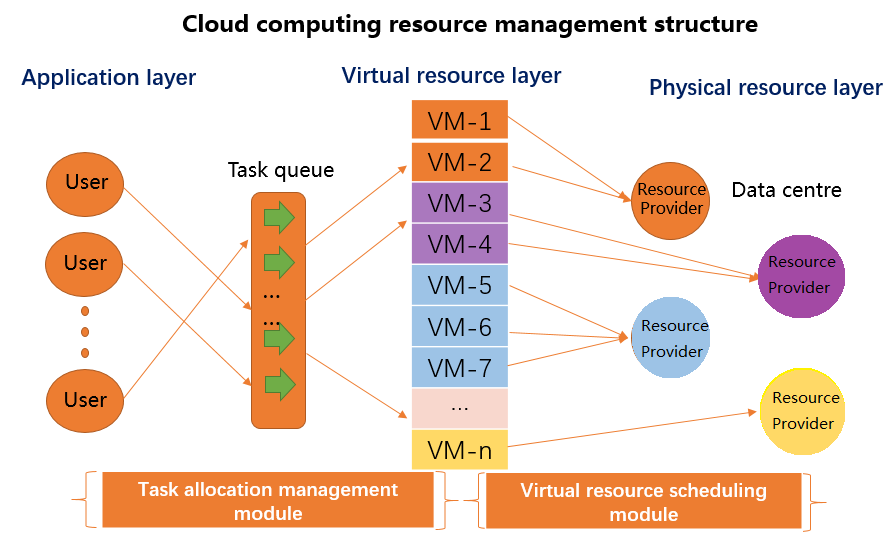}
\caption{Cloud computing resource management framework}
\label{Fig1}
\end{figure}

This paper revisits the issue of QoS management in cloud computing by proposing models for both task allocation to virtual machines and virtual machine migration across physical resources as shown by the model depicted in figure \ref{Fig1}. This model has two parts: a  task allocation module  located  between application layer and virtual resource layer and a virtual machine migration module which is between virtual resource layer and physical resource layer.

As depicted by the figure, the cloud computing resource management framework is made up of the following layers:
\begin{enumerate}
\item A Physical resource layer, which is composed of the data centres. This is layer hosts different hardware resources in the form of host machines.
\item A virtual resource layer sits on top of the physical resource layer. Within this layer, the physical hosts are broken into virtual hosts for better resource management. 
\item An application layer is layered above the virtual resource layer to provide different services to the users. These include Software  as a Service (SaaS),  Platform  as a Service (PaaS) and in special cases Infrastructure as a Service (IaaS).
\end{enumerate}

When considering a service perspective, the framework in figure~\ref{Fig1} can be presented as a two-layer architecture including:
\begin{enumerate}
\item A task allocation management module which enables cost effective way allocation of virtual resources to users.  
\item A virtual resource scheduling module, where a mapping between physical resources and the virtual machines is made. We assume In this paper that each host provides at least one virtual machine(VM). 
\end{enumerate}

The focus of this paper lies on the management of task allocation to virtual machines and the migration of virtual machines to physical machines. Note that while the task allocation to virtual machine is a federation agnostic process, the migration of virtual machines to physical resources is a federation-aware process. Federation awareness allows the resource providers freedom to either behave cooperatively by leasing their resources to a common pool when they are not needed or competitively and pull back the resources for their own use when needed. The main contributions of this paper are as follows:
\begin{enumerate}
\item	 A multi-QoS based resource allocation management model in cloud computing environment. We performed QoS-based modelling of both task allocation and virtual machine migration in cloud computing environments. These models are designed to maximize the overall benefits of the system while adhering to QoS constraints.
\item	 A multi-QoS based method of task scheduling with Differential Evolution (DE). We propose the DE based binding policy (DEBBP) as a solution for task assignment and use it to implement a new binding strategy for the popular CloudSim simulator. The effectiveness of the DE algorithm for multi-QoS target task scheduling is proved by comparing it with the conventional binding policy of cloud-sim and the MaxMin algorithm-based binding policy.
\item	 A VM Migration Necessity-based Dynamic Scheduling Algorithm (VMMNDSA). In order to reduce the energy consumption of the overall virtual machine migration while keeping the level of SLA (QoS performance), we proposed an efficient virtual machine migration algorithm that takes into account load capacity assessment and target host selection to solve and optimize the VMs migration problem in cloud computing environment. The efficiency of our proposed model was proved by comparing it with the Dvfs algorithms and LrMc algorithms.
\item  Cloud federation modelling and evaluation which considered virtual machine migrations under both cooperation and competition federation modes. We then extended the CloudSim framework by incorporating these models into it. 
\end{enumerate}	 

The work presented in this paper is aimed at complementing the models proposed in~\cite{cyberhealth,kiosks,hframe,priority,hl7cda,fiware} with the expectation of developing efficient healthcare infrastructures in the rural and urban areas of the developing countries 
where cloud resources might be sufficient in some countries and limited in other countries. A simple illustrative scenario in resource constrained countries consists of alternating between cooperative cloud computing when cloud providers have enough resources for their clients and an excess to borrow to the cooperation and competitive cloud computing when cloud providers have only limited resources for their users' applications.

\section*{Related Works}
At present, researchers have done a lot of research works in the field of task scheduling and resource management. In terms of the task scheduling, different computing resource has different specification and each task also has its own request, which means finding the best solution between task requests and computing resources is a NP hard problem. Based on this, a large number of heuristic algorithms has been proposed and used for the approximate solution of the above optimal matching problem. In many application task scheduling strategies, Min-Min method \cite{He} and Max-Min method \cite{Chauhan} are often used as benchmark for evaluating the performance of other scheduling strategies. These two principles are similar and are arguably the most representative classical heuristic algorithms. For specific task scheduling problems, due to the superiority of survivability, some intelligent optimization algorithms, such as the genetic algorithm \cite{Gao} for practical scientific workflow, are also used to approximate the global optimal solution of task scheduling problem. For the computationally intensive application task, Daniel \cite{Daniel} proposed a scheduling method that can handle multiple application tasks in heterogeneous platform at the same time. These applications could either be single and independent tasks or multiple tasks (bag of tasks). A mixed-game model designed for task-type preference and resource-service capability was proposed by Li \cite{LI}. The work focused on QoS optimization by usig the multi-services-based evolutionary game scheduling algorithm. Considering QoS constraints of task scheduling, Oprescu \cite{Oprescu} proposed a budget constrained task scheduling scheduler, which is able to deal with multiple task bags under different minimum completion time constraints, maximum budget constraints and different performance parameters.

Virtual machine migration can also be considered a form of resource allocation and Marzolla \cite{Marzolla} proposed a conversation-based virtual machine integration algorithm. The algorithm exchanges information between computing nodes through a simple conversation protocol and integrates virtual machines based on the information obtained. They aimed at maximizing the average resource utilization of computing nodes. Perumal and Subbiah \cite{Perumal} took memory utilization into account when integrating running virtual machines in cloud data centres, and proposed an algorithm based on Optimal Virtual Machine Placement (OVMP). The algorithm first finds all compute nodes whose memory utilization are lower than the set threshold then consolidates the virtual machines running on these compute nodes together. Upon successful consolidation, idle compute nodes are shut down. Shi \cite{Shi} considered the number of migrations in the virtual machine integration problem and established a virtual machine integration model that minimizes the number of compute nodes and virtual machine migration times. Beloglazov \cite{Beloglazov2} proposed the Double Thresholds VM Selection (DTVS) algorithm to integrate virtual machines running in a cloud data centre. A summary of some research works on QoS in cloud computing and federated cloud computing is presented in Table~\ref{rel}.

\begin{table}[!ht]
\caption{Summary of related works.}
\begin {center} 
\label{rel}
\begin{tabular}{|p{0.8 cm}|p{2.14 cm}|p{9.60 cm}|}
\hline
Work                                                              & Type          & Summary                                                                                                                                                                                                                                                                                                                                                                                                                                                                                                                                                                                                                \\ \hline
\cite{Yu}                                 & unfederated & \begin{tabular}[c]{@{}l@{}}They devised a novel online scheduling framework\\ (MDTC-HB) consisting of an offline scheduling \\phase and an online rescheduling phase, which can \\provide rapid response to dynamic VM migrations\\ while comparing with the existing approaches.\end{tabular}                                                                                                                                                                                                                                                                                                                         \\ \hline
\cite{Patel}           & unfederated & \begin{tabular}[c]{@{}l@{}}This work proposed an Energy-Aware VM \\ Migration algorithm approach to balance the load \\on PM. VM migration due to temporary peak load\\ is one of the critical issues in cloud, which highly \\impact the performance of the cloud services. To \\avoid these unnecessary migrations, the approach\\ uses a prediction approach which predicts the\\ load on the PM before triggering the migrations.\end{tabular} \\ \hline
\cite{Mandal}         & unfederated & \begin{tabular}[c]{@{}l@{}}They propose a strategy to determine appropriate \\migration bandwidth and number of pre-copy \\iterations, and perform  numerical experiments in \\multiple cloud environments with  large number of \\migration requests. Coupled with maximum and \\minimum-bandwidth provisioning strategies while\\ using  an order of magnitude less bandwidth than\\ maximum bandwidth strategy. It also achieves\\ significantlylower migration duration than \\ minimum-bandwidth scheme.\end{tabular}                                                                                                  \\ \hline
\cite{Vijin}           & federated     & \begin{tabular}[c]{@{}l@{}}An innovative economic sharing model is used to \\sharing  capacity in a federationof IaaS cloud \\providers, this  can be done by using interaction \\among cloud providers as a repeated game of VM\\ that can identify  the all unused capacity in the \\spot market. Their work more focuses on the \\conditions of security, not the design of the \\federation itself.\end{tabular}                                                                                                                                                                                      \\ \hline
\cite{Habibi}         & federated     & \begin{tabular}[c]{@{}l@{}}They look at statistical multiplexing and server \\ consolidation as such a strength and examine the \\use of the coefficient of variation and other\\ related statistical metrics as objective functions.\\ The results show that their algorithm based on \\the Late Acceptance Hill Climbing method \\outperformed others.\end{tabular}                                                                                                                                                                                                                                                  \\ \hline
\cite{Moghaddam} & federated     & \begin{tabular}[c]{@{}l@{}}In this paper, a federated policy-based resource \\ classification model has been presented to \\classify and manage security levels in clouds and \\to provide efficient mapping between access \\requests and defined  policies of each cloud node. \\The reliability and efficiency of this policy-based \\ classification schema have been evaluated by \\performance, security and competitive analysis.\end{tabular}                                                                                                                                                      \\ \hline
\end{tabular}
\end{center}
\end{table}

\section*{Multi-QoS task scheduling model}

This section  introduces the design of task allocation model by starting by presenting a QoS formulation of the task allocation module and thereafter presenting the architecture of the task allocation process which mimics its underlying algorithm. Finally, a differentiated evolution (DE) algorithm expressing  how the binding policy is implemented by the process is presented.

\subsubsection*{QoS formulation of the task allocation module}
QoS in cloud services can be measured with different attributes, some of which are summarized in Table~\ref{tab2}.

\begin{table} [!ht]
\caption{QoS attributes.} \label{tab2}
\begin{center}
\begin{tabular}{|l|l|}

\hline
QoS attributes			& Description\\ 
\hline
Execution price ($P_{ex}$)		& The cost required for the cloud service \\&provider to perform the task.\\

Execution time ($T_{ex}$)		& The time it takes for the cloud service  \\&provider to perform the task.\\

Reliability ($R_{el}$)			& The probability of a cloud service running \\&normally.\\

Availability ($A_{va}$)			& The probability that a cloud service can be\\ &successfully accessed.\\

Security ($S_{ec}$)				& The security level of the cloud service.\\
\hline
\end{tabular}
  \end{center}
\end{table}

These attributes have been selected as key parameters of the QoS model proposed in this paper and Table~\ref{symb} summarizes the symbols used by the model. These parameters are:

\begin{table*} [h]
\caption{Symbols meaning} \label{symb}
\begin{center}
\begin{tabular}{|l|p{9.0cm}|}
\hline
Symbol          			& Meaning                                                                                                                                                       \\ \hline
S        				& expresses a mapping between tasks and virtual machines upon completion \\
                           &of the task scheduling process. \\ 
$\alpha(i,j)$        		& boolean value that reveals  if there is a match  between task  $i$ and virtual machine  $j$.     \\ 
F(S)’           			& the influence vector of execution time and execution price between tasks and VMs                                                                       \\ 
$(1-\frac{T_{ex}(i,j)}{D_{ead}})$  & is the benefit related to the time  saved upon completion of the  task.                                                                                          \\ 
$(1-\frac{P_{ex}(i,j)}{B_{udg}})$  & is the benefit related to the cost savings upon completion of the task.                                                                                          \\ 
RAS(S)          			& \begin{tabular}[c]{@{}l@{}} is the dependability score of the system S.					
\end{tabular}  \\ 
$R_{el}(j), A_{va}(j), S_{ec}(j), m_j$ 	& \begin{tabular}[c]{@{}l@{}} $R_{el}(j), A_{va}(j), S_{ec}(j)$ respectively express the reliability,\\ availability and security of the VM j. $m_j$ expresses  \\the number of tasks that the j-th VM are executed.\end{tabular} \\ 
F(S)            			& is the final fitness function which represents the score of the mapping S                                                                                                \\ 
 $\alpha$, $\beta$             & \begin{tabular}[c]{@{}l@{}}the weight factors to respectively adjust the importance\\ of execution time and price\end{tabular}                                              \\ \hline
\end{tabular}
 \end{center}
\end{table*}

i)  the scheduling deadline ii) the scheduling budget and iii) a provider dependability parameter that combines reliability, availability and security. 

While the scheduling deadline defines the longest/maximum  time taken by the tasks to be executed on all virtual 
A task scheduling model named Multi-QoS-Target Scheduling Model (M-QoS-TSM) is proposed in this paper as solution of the task allocation management problem. As proposed in this paper, the M-QoS-TSM model focuses on three QoS target parameters: 
machines, the  scheduling budget expresses the total cost of completing the execution of all tasks on all the virtual machines in a specified target  environment. The reliability, availability and security are attributes which are specific to different resource providers, 
hence also the provider dependability. They may vary from a resource provider to another and may be a defining factor for users to prefer one provider compared to the other.  
Suppose the cloud computing virtual resource is consist of virtual machines, which can be represented as a virtual machine set 
$VM= \{ i, i \in 1,\dots N \}$ and  the application tasks which are going to be scheduled can be represented as a task set $T = \{ j, j \in  1, \dots n \}$. 

From a user's perspective, the cloud computing system should allow them to complete their tasks on time while making savings on the execution costs.  These two objectives can be expressed  by a multi-objective optimisation model aiming at maximising both execution time and execution price subject to constraints on deadlines and budget as expressed below:

\begin{equation} \label{prb1}
\min_{i,j} 
\left \{
\begin{array} {lr}
\max_{j \in N} \sum\limits_{i=1}^{n} T_{ex}(i,j)  * \alpha(i,j)			&  \mbox{    Execution time }	 \\
\sum\limits_{i=1}^{n} \sum\limits_{j=1}^{N} P_{ex}(i,j) 	* \alpha(i,j)	& \mbox{     Execution price } 
\end{array}\right. 
\end{equation}
$$
 \mbox{                                    subject to } 
$$
$$
\left \{
\begin{array} {llr}
\max_{j \in N} \sum\limits_{i=1}^{n} T_{ex}(i,j))		& \le D_{ead}  	&  \mbox{     (\ref{prb1}.a)}	 \\
\sum\limits_{i=1}^{n} \sum\limits_{j=1}^{N} P_{ex}(i,j) 		& \le B_{udg}  	& \mbox{     (\ref{prb1}.b)} 
\end{array}\right.
$$
Where 
$$
\alpha(i,j) =
\left \{
\begin{array} {ll}
1 	&  \mbox{    if task $i$ matches VM $j$ }	 \\
0 	& \mbox{     Otherwise } 
\end{array}\right. 
$$
and $D_{ead}$ is the execution deadline of all tasks on virtual machines, $B_{udg}$ is the total budget allocated to the execution of tasks on virtual machines,  $T_{ex}(i,j)$   describes the execution time of the task $T_{i}$ , ($i \in n$) on the  
$VM_{j}$,  ($j \in N$) and $P_{ex}(i,j)$ is the price associated with the execution of task $T_{i}$ by the virtual machine $VM_{j}$.

Thus, a comprehensive QoS performance metric can be found by integrating the two parameters above into a benefit (fitness) function which is defined by: 

\begin{equation} \label{prb2}
F(S)' = 
\left \{
\begin{array} {lll}
\alpha(1-\frac{T_{ex}(i,j)}{D_{ead}})+ \beta(1-\frac{P_{ex}(i,j)}{B_{udg}}),  \\	  \mbox{  if } 	\alpha(i,j) =1 \; \land  \; T_{ex}(i,j)\leq D_{ead}\; \land \; \\P_{ex} (i,j)\leq B_{udg}  \\ \\
0,  										\\	 \mbox{  if }   \alpha(i,j) =0 \; \lor \; T_{ex}(i,j)> D_{ead} \; \lor \; \\P_{ex} (i,j)> B_{udg}
\end{array}\right. 
\end{equation}

\begin{equation}
\begin{split}
RAS(S)&= \sum \limits_{j=1}^{m_j} ((R_{el}(j)+A_{va}(j)+S_{ec}(j))* m_j
\end{split}
\end{equation}

\begin{equation}\label{eq5}
  F(S)=\begin{cases}
     F(S)' * RAS(S), & F(x_{i})' \neq 0 \\
     0, & F(x_{i})' = 0
  \end{cases}
\end{equation}

\subsection*{The  architecture of the task allocation  process}
\begin{figure}[h]
\centering
\includegraphics[scale=0.5]{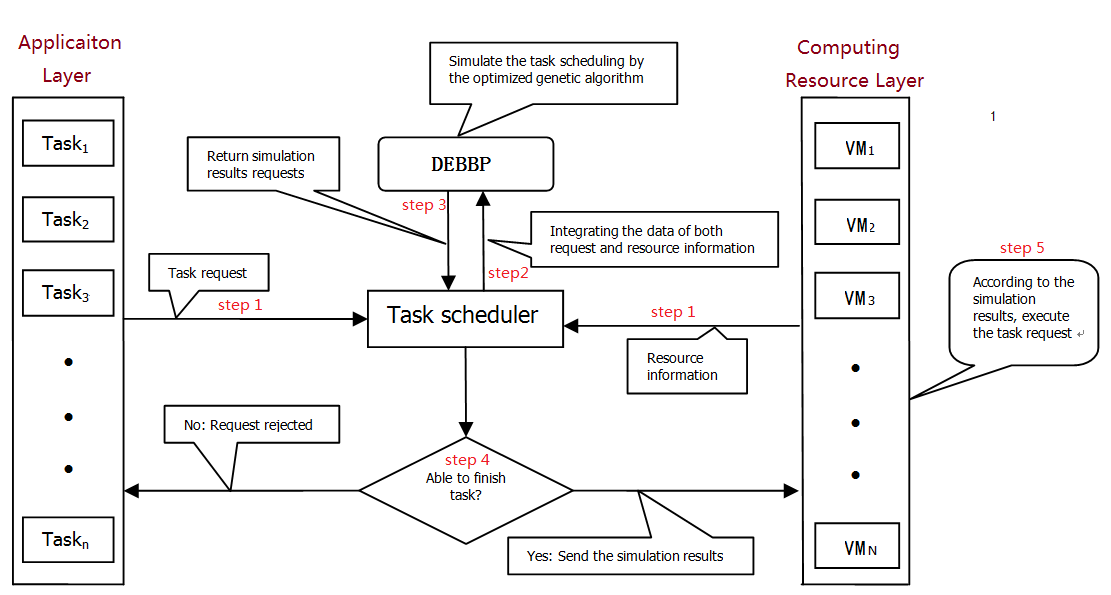}
\caption{The architecture of M-QoS-TSM}
\label{Fig2}
\end{figure}

Figure \ref{Fig2} gives the architecture of the proposed M-QoS-TSM model. Its implementation processes are as follows: at the first step users submit the application tasks with associated QoS constraints to the task scheduler, and at the same time, VMs send their resource information to the task scheduler. At the second step, the scheduler generates the simulation results based on DEBBP policy. It then returns the simulation results to task scheduler in the third step. At the fourth step, the scheduler judges the returned information, if the VMs in the cloud system are able to finish the users’ task request, then the simulation results are accepted and the tasks allocated to the obtained VMs. If otherwise, the scheduler rejects the request and informs the users of the rejection due to the resource power limitations. At the fifth and final step, the VMs execute the allocated tasks according to the received simulation result.

\subsection*{The DE Based Binding Policy}
As mentioned before, task scheduling in cloud computing environment is a NP hard problem. It is very hard to find the best solution while the quantity of participants is big. The usual way is to apply various intelligent optimization algorithms to approach its optimal solution as the satisfactory solution. DE algorithm \cite{Sun} is one of these algorithms to get the approximate optimal solution. It generally requires the parameters of three main steps, population size (NP), a parameter (F) to control the mutation, as well as the crossover probability (CR). 

Initialing population: creating a population X consists of individuals $x_{i}$ = {$x_{i,1}, x_{i,2}, …, x_{i,D}$} T, where i = 1,…,NP, and {$X_{i}(0) | x^{L}_{i,1} \leq x_{i,1}(0) \leq x^{U}_{i,1}; i= 1,2,…, NP; j=1,2…,D$}, where $x_{i}$(0) is the ith individual and j means j-th dimensionality, and the way to calculate xi,j(0) is described by
\begin{equation}\label{eq6}
	x_{i,j}(0) = x^{L}_{i,j} + rand(0-1)*(x{U}_{i,j} - x{L}_{i,j})
\end{equation}
where $x^{L}_{i,j}$, $x^{U}_{i,j}$ respectively express the lower bound and upper bound of j-th dimensionality, rand(0-1) means a random number between 0 and 1. In DEBBP, individuals $x_{i}$ indicates one match result between VMs and tasks, and D is the number of tasks.

Mutation: The DE algorithm implements individual mutation through a differential strategy. The common difference strategy is to randomly select two different individuals: $x_{r2}$ and $x_{r3}$ in the population, then use a scaling factor (F) to scale the difference between $x_{r2}$and $x_{r3}$. Than we can generate the mutant vector by adding another random population $x_{r1}$ by
\begin{equation}\label{eq6}
	V^{g+1}_{i} = x^{g}_{r1} + F(x^{g}_{r2} - x^{g}_{r3}), 
\end{equation}
where r1, r2 and r3 are three random number in [1, NP], the F is a certain constant, and g indicates the g-th generation.

Crossover: The purpose of crossover operation is to randomly select individuals. The method of crossover operation can be described by
\begin{equation}\label{eq7}
\begin{split}
	U^{g+1}_{i,j} &= V^{g+1}_{i,j} if rand(0,1) \leq CR, j=1,...,D \\ 
       &=x^{g}_{i,j} otherwise,
\end{split}
\end{equation}
where CR is the cross probability. The crossover operation refers to optimization process. If $F(U^{g+1}_{i})$ < $F(x^{g}_{i})$ then $x^{g}_{i} = U^{g+1}_{i}$, where F(·) is the fitness function (5).
Description in pseudo-code of applied DE is presented in Algorithm 1.

\begin{algorithm}[h]
\SetKwInOut{Input}{input}\SetKwInOut{Output}{output}

\Input{ NP: population size, F: a parameter to control the mutation, CR: crossover probability, fitness function: f(·) - equation (5), a random population: $x^{G}_{i}$, generation: the times of generation repeat}   
\Output{Optimal solution (task scheduling results): individuals from last population with best fitness} 
int t = 0;  \\
while t $\leq$ generation do \\
foreach vector $x^{G}_{i}$ from population X
do Generate mutant vector viG according to (8), Crossover vectors according to (9) \\
if $F^{g+1}_{i} < F^{g}_{i}$  \\
then $x^{g}_{i} = U^{g+1}_{i}$ \\
\textbf{end} \\
\textbf{end} \\
t = t + 1 \\
\textbf{return} Individuals of the last population with best fitness
\caption{Mutation. \label{Gre:Ass2}} 

\end{algorithm}

\section*{Virtual Machine Migration Model}
The VM migration model is mainly composed of two layers: physical resource layer and virtual resource layer which are shown in figure 1. The virtual resource layer provides the operating carrier for the application layer, while being itself built upon the physical resource layer, that is the virtual server. In terms of quantity, a virtual machine can hold more than one application. The mapping relationship between application systems and virtual machines is therefore many-to-one. The physical resource layer, which provides hardware resources for the virtual resource layer, is mainly composed of various types of hardware servers in the data centre. The mapping between virtual machines and physical machine is also a many-to-one, which means one physical machine can host multiple virtual machines. The access connection requirements of application services in this model are dynamic, hence the cloud resource management platform can scale (increase or decrease) the number of virtual machines to meet the system to meet application service demands. This is achieved by moving (migrating) VMs across hosts. This VM migration process affects the overall utilization level of physical servers, the balance level and energy consumption in data centre.

\subsection*{The cloud federation model}
Typically, a  federated model for cloud computing includes different cloud providers collaborating by i) sharing their resources while having each of them remaining an independent autonomous entity by keeping keeping thick walls in between them by and ii) having the applications running in this cloud of clouds while being unaware of location due to virtual local networks being designed and implemented to enable  the inter-application components to communicate and iii)  having cloud providers differentiate from
each other in terms of cost and trust level. 

\begin{figure}[h]
\centering
\includegraphics[scale=0.35]{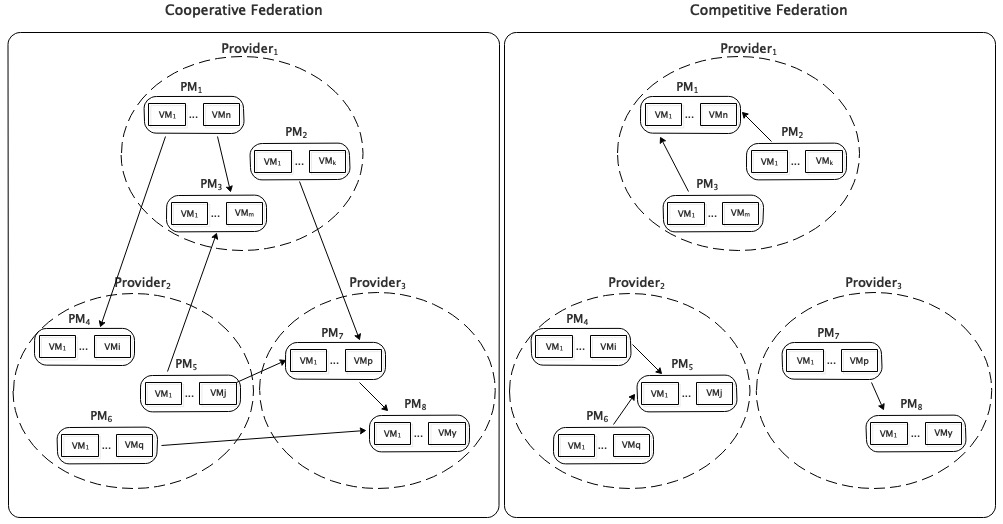}
\caption{Cloud computing resource management framework}
\label{federation}
\end{figure}

When considering a federated cloud environment, the virtual machines allocated to the users' tasks can be migrated either to physical resources of the users' current cloud provider or to physical resources of different cloud providers. Such an allocation of virtual resources to physical resources can lead to a cooperative model when users's virtual machines can be migrated anywhere or a competitive model when users' virtual machines can only be migrated to their providers' physical machines as expressed by Figure~\ref{federation}.

 In our model, the resources which have been availed by a PM $k$ of provoder $x$ , which are expressed by $R_{pn}(k_{x})$ while demand for resources by a VM $j$ (it used to belong to provider $y$) during migration are expressed by $D_{vm}(j_{y})$.  The federated cloud computing  problem consists of finding for 
each VM $j$ in distress, a mapping to a physical resource PM $k$ that maximises a utility function $D(j,k)$ as defined below:
\begin{equation} \label{diff}
\begin{array}{ll}
\max D(j,k) & = P(j_{y},k_{x}) * (R_{pn}(k) - D_{vn}(j))   \\
subject & \mbox{ to } 
\end{array}
\end{equation}
$$
\left \{
\begin{array} {llr}
R_{pn} (j) \ge D_{vn} (k)  & \forall j \in {\mathcal V}, k \in {\mathcal P}    &\mbox{     (\ref{diff}.a)}\\
\forall j \in {\mathcal V}, k \in {\mathcal P}  & \rightarrow P(j_{y},k_{x}) \in \{0,1\}  &\mbox{     (\ref{diff}.b)}
\end{array}\right.
$$
Note that as expressed by equation $\ref{diff}.b$,   $P(j_{y},k_{x})$ is a binary parameter used in the model to differentiate between cooperative and competitive cloud computing as expressed below:
\begin{equation} \label{pij}
P(j_{y},k_{x}) = \left \{
\begin{array} {llr}
1   &  x\not=y   &  \mbox{   Cooperative cloud computing}\\
0   &  x=y & \mbox{    Competitive cloud computing}
\end{array}\right.
\end{equation}

Note that as expressed by equation~\ref{pij},  $P(j_{y},k_{x})$ is used in the model to enable all participant providers to be elected for a VM migration under cooperative cloud computing and discard providers from participating in a VM migration under competitive cloud computing when the VMs don't belong to their clients. 

\subsection*{VM Migration Necessity-based Dynamic Scheduling Algorithm based on migration technology}

Given the high overhead required of VM migration, unreasonable migrations may further reduce the performance of cloud computing data centre. Therefore, our model will mainly focus on logically determining when and where to migrate virtual machines, based on dynamic scheduling model of cloud resources, as well as virtual machine migration technology \cite{Kaur}. We proposed a VM Migration Necessity-based Dynamic Scheduling Algorithm (VMMNDSA), which considers the load on the physical node, virtual machine migration loss assessment and target host positioning, to achieve efficient load balancing through reasonable resource scheduling. The architecture of VMMNDSA is shown in Figure \ref{fig3}:

\begin{figure}[h]
\centering
\includegraphics[scale=0.5]{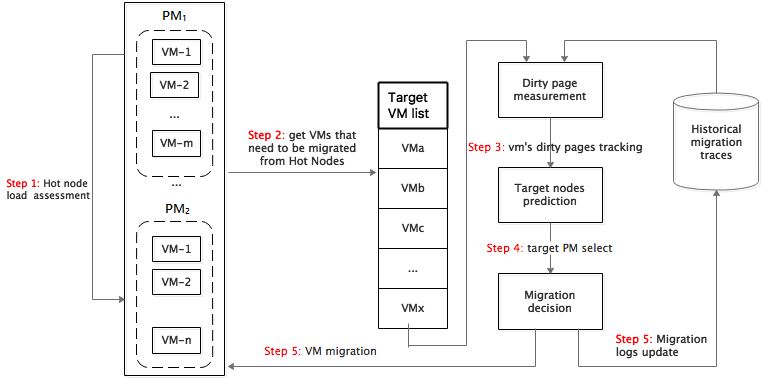}
\caption{The architecture of VMMNDSA Based on Migration Technology}
\label{fig3}
\end{figure}

\begin{enumerate}
\item~Hot node load assessment\\
We compared each physical node (PM) with $T_{max}$ and $T_{min}$, to get the PMs, that PM\textgreater $T_{max}$ or PM\textless $T_{min}$, as the hot nodes.

\item~Virtual Machine Migration Loss Assessment\\
This assessment is used to determine a suitable virtual machine as an object to be migrated from an overloaded physical node to reduce the load of the physical node. The content of the VM which is going to be migrated mainly includes the migration of running status and data resources, as well as memory migration of the migration data resources. Therefore, the cost of migration needs to be considered when determining the target VM.

The definition of variable migration loss ratio: cm = $(\alpha$'*$c_{j}$+ $\beta$'*$b_{j})/m_{j}$, where $c_{j}$ is the CPU utilization of the j-th virtual machine on the overloaded physical node, $b_{i}$ is the network bandwidth utilization of the j-th virtual machine on the overloaded physical node, and $m_{j}$ is the memory utilization of the j-th virtual machine on the overloaded physical node, and $\alpha$', $\beta$' are the weight factor to adjust the importance of the CPU and bandwidth. According to the selecting sort method to select the virtual machine with the largest migration loss ratio as the object to be migrated, the greater the CPU resource utilization $c_{i}$ and the bandwidth $b_{i}$ are, the more the computing resource is consumed, and the more the computational resource load of the physical node can be relieved. Meantime, the smaller the memory resource utilization is, the smaller the migration overhead is due to less data need to be deal with. Therefore, the migration loss ratio can well locate which VM to be migrated of the overloaded physical node.
\item~Dirty pages tracking\\
Once the VM begin to be migrated, the VMs are not able to work until the migration is finished, and their memory will be transferred in several times of iteration from the hot node to the destination compute node, that is, the transferred memory page size is the portion that was modified, and the modified page is called "Dirty Pages." Finally, the hot node stops the VM operation and transmits the “dirty page” generated by the last iteration to the destination compute node. The dirty pages management is used for managing the transformation process of Dirty Pages.
\item~Target node (PM) selection\\
The positioning of target physical node is determined based on the best matching method. The variable Divm is defined to represent the target virtual machine demand vector, the defined variable $\Delta$D means the gap of different resource vectors between virtual machine demand vector and all the candidate physical nodes.

The performance of a physical node depends on how much resource it left. This paper mainly considers the performance of three resources: CPU, memory, and network bandwidth. So the detail of equation $\ref{diff}$ can be represented by the $R_{pn}(k)$ = [$r_{kcpu}$*$u_{kcpu}$, $r_{kmem}$*$u_{kmem}$, $r_{kbw}$*$r_{kbw}$] of physical node k, $r_{kcpu}$ represents the total resource of CPU and $u_{kcpu}$ represents the usage of CPU resources, $r_{kmem}$ represents the total resource of memory and $u_{kmem}$ represents the usage of memory resources, $r_{kbw}$ represents the total resource of bandwidth and $r_{kbw}$ represents the usage of network bandwidth resources. 

The virtual machine demand vector is $D_{vm}(j)$ = [$D_{jcpu}$, $D_{jmem}$, $D_{jbw}$], $D_{jcpu}$, $D_{jmem}$ and $D_{jbw}$ respectively indicates the CPU demand vector, memory demand vector, and bandwidth demand vector of the virtual machine for the j-th physical node. The formula is shown as follows:\\
$D_{jcpu}$ = the amount of CPU resources required by the virtual machine /CPU total resources. \\
$D_{jmem}$ = amount of memory resources required by the virtual machine/total memory resources.\\
$D_{jbw}$ = amount of bandwidth resources required by the virtual machine /total bandwidth resources.

Then according to equation  $\ref{diff}$, the $max D(j,k)$ shows that the $PM_{k}$ will be the target node of $VM_{j}$.

\item~Actual VM migration and migration logs updates\\
\end{enumerate}

\section*{Experiment results}
CloudSim was released by Melbourne University and Gridbus project group in 2009, as a toolkit for simulating cloud computing environment \cite{Rodrigo}. It is based on the existing Java based discrete event simulation package in GridSim. The operating environment of simulation experiment in this paper is as follows: CPU: Intel Pentium dual core P6000, clocked at 1.87 GHz; Memory :8.0 GB; and HDD: 1 TB.

\subsection*{Implementation of the Proposed DEBBP for experiment results of task allocation management model}
In the CloudSim environment, 8 computing nodes were used, each with different attribute parameters. The performance configuration and computing power were modeled after the Amazon EC2 on-demand instance pricing model and showed on Table \ref{tab444}. Then we simulated job A and job B using DEBBP, the conventional binding policy and the MaxMin based binding policy (MMBBP) to allocate cloudlets to VMs. Each job has 20 cloudlets which needed to be assigned to VMs. The information of these two jobs are shown on Table \ref{tab555}, and the weight factors are set as: $\alpha=\beta$  ($\alpha+\beta$=1), deadline and budget are respectively 100 and 1800. More information about the chosen resource allocation scheme can be found in the article \cite{Ian}.

\begin{table}[!ht]
\caption{VM List of task allocation management model}
\label{tab444}
\begin{center}
\begin{tabular}{lllllllll}
\hline
VM ID              & VM0 & VM1 & VM2 & VM3 & VM4 & VM5 & VM6 & VM7 \\ 
Pes                & 1   & 1   & 1   & 1   & 1   & 1   & 1   & 1   \\ 
Mips               & 101 & 410 & 213 & 54  & 55  & 363 & 70  & 118 \\ 
Price              & 5   & 20  & 10  & 5   & 5   & 7   & 6   & 11  \\ 
Availability (AVA) & 40  & 10  & 20  & 20  & 30  & 50  & 10  & 30  \\ 
Reliability (RE)   & 10  & 10  & 10  & 60  & 20  & 40  & 10  & 20  \\ 
Security (SE)      & 20  & 40  & 10  & 40  & 10  & 60  & 50  & 30  \\ \hline
\end{tabular}	
\end{center}
\end{table}

\begin{table}[!ht]
\caption{VM List of task allocation management model}
\label{tab555}
\begin{center}
\begin{tabular}{lll}
\hline
Cloudlet ID & Job A (Length) & Job B (Length) \\
\hline
0        & 2020       & 930           \\
1        & 700        & 110           \\
2        & 170        & 310           \\
3        & 100        & 1300          \\
4        & 440        & 840           \\
5        & 620        & 120           \\
6        & 710        & 310           \\
7        & 660        & 60            \\
8        & 820        & 2220          \\
9        & 820        & 290           \\
10       & 1100       & 600           \\
11       & 300        & 820           \\
12       & 300        & 1150          \\
13       & 40         & 670           \\
14       & 1750       & 930           \\
15       & 460        & 260           \\
16       & 330        & 300           \\
17       & 110        & 410           \\
18       & 550        & 500           \\
19       & 900        & 90   \\
\hline        
\end{tabular}
\end{center}
\end{table}

For the DE algorithm used in this paper, the operational parameters are adapted from those used by several other related publications and are as follows: the population size was set to 100, maximum and minimum value were set to 8 and -8 respectively; the mutation zoom was set to 0.5, the crossover to 0.1 and the maximum iteration set to 2000.

\begin{figure}[!ht]

\begin{center}
\begin{subfigure}{0.45\textwidth}
		\includegraphics[width=\textwidth]{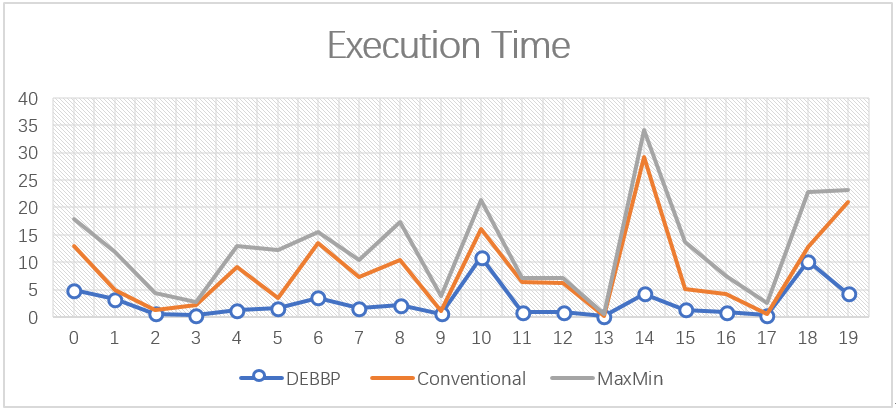}
		\caption{Job A}
		\label{fig4a}
	\end{subfigure}
	\begin{subfigure}{0.45\textwidth}
		\includegraphics[width=\textwidth]{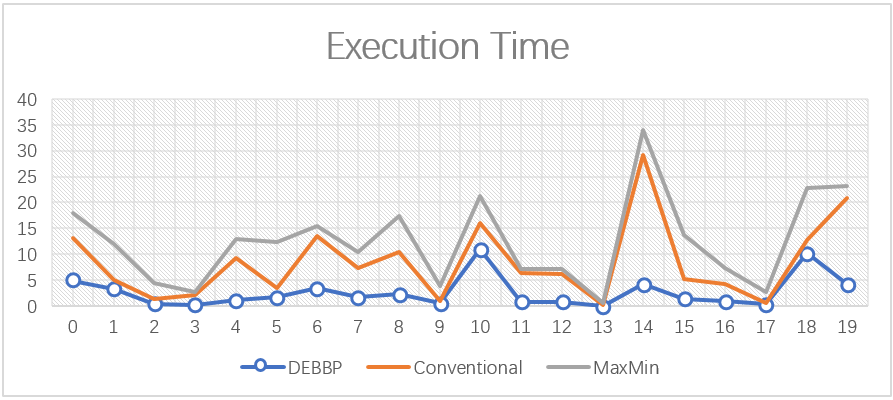}
		\caption{Job B}
		\label{fig4b}	
\end{subfigure}
\caption{The experiment results of execution time}
\label{fig555}
\end{center}
\end{figure}

\begin{figure}[!ht]
\begin{center}
\begin{subfigure}{0.45\textwidth}
		\includegraphics[width=\textwidth]{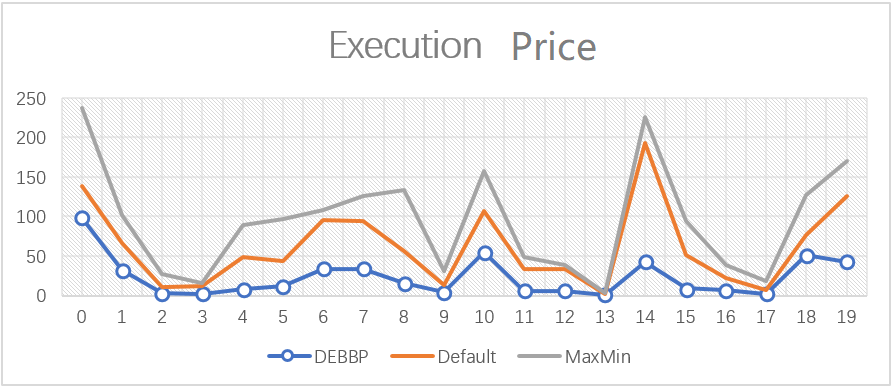}
		\caption{Job A}
		\label{fig5a}
	\end{subfigure}
	\begin{subfigure}{0.45\textwidth}
		\includegraphics[width=\textwidth]{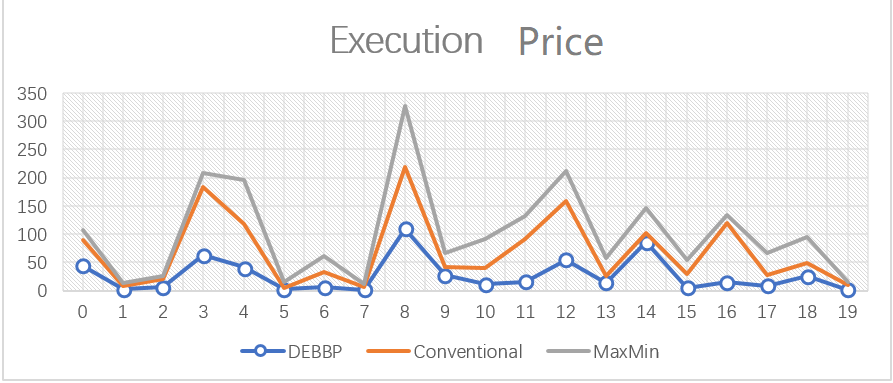}
		\caption{Job B}
		\label{fig5b}	
	\end{subfigure}
\caption{The experiment results of execution Price}
\end{center}
\end{figure}

\begin{figure}[!ht]
\begin{center}
\begin{subfigure}{0.45\textwidth}
		\includegraphics[width=\textwidth]{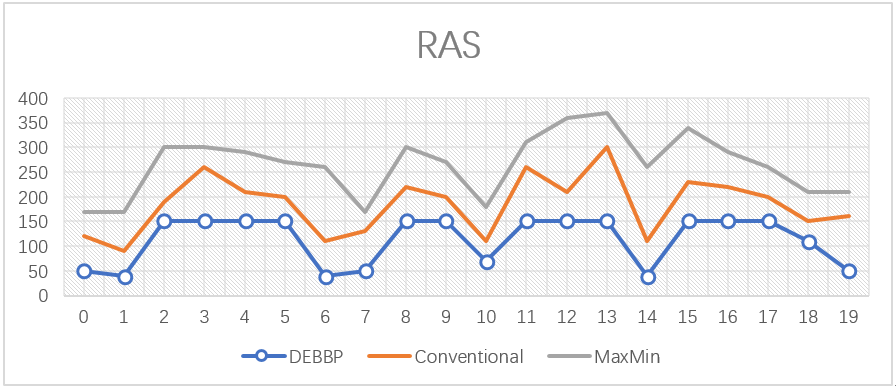}
		\caption{Job A}
		\label{fig6a}
	\end{subfigure}
	\begin{subfigure}{0.45\textwidth}
		\includegraphics[width=\textwidth]{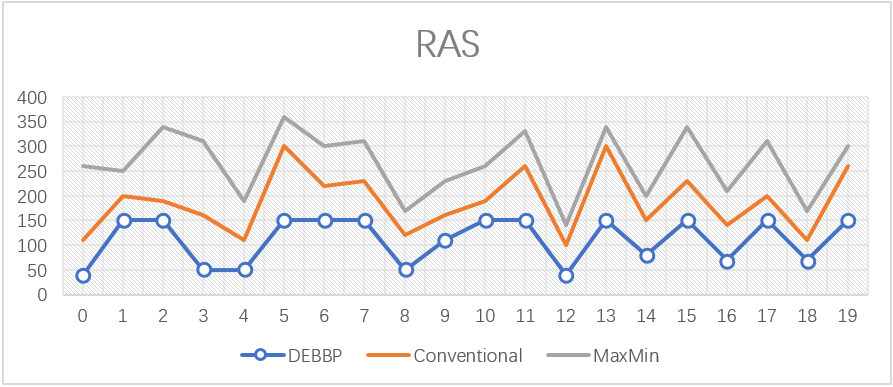}
		\caption{Job B}
		\label{fig6b}	
	\end{subfigure}
\caption{The experiment results of RAS}
\end{center}
\end{figure}

\begin{figure}[!ht]
\begin{center}
\includegraphics[width=6 cm]{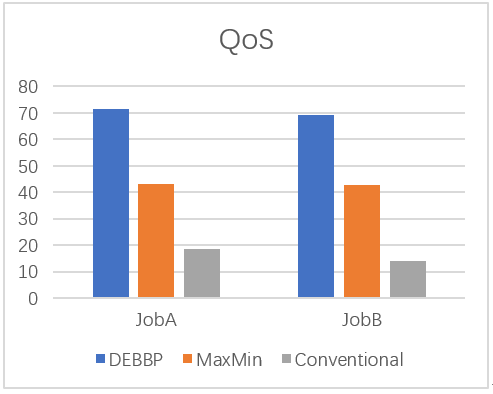}
\caption{The comprehensive QoS perfermance}
\label{fig7}
\end{center}
\end{figure}

The simulation results are presented on Tables \ref{tab6} and \ref{tab7}; while information on execution times are shown in figure \ref{fig4a} and figure \ref{fig4b}. Information on execution prices are shown in figure \ref{fig5a} and figure \ref{fig5b}, while those of RAS are shown in figure \ref{fig6a} and figure \ref{fig6b}. Finally, a plot of the QoS is shown in figure \ref{fig7}. Cloudlets are also mapped to the VMs sequentially, that is cloudlet ID 0 to VM ID 0, cloudlet ID 1 to VM ID 1, cloudlet ID 2 to VM ID 2, etc. DEBBP however allocates cloudlets to VMs differently, with cloudlet ID 0 being allocated VM ID 5, cloudlet ID 1 to VM ID 2, cloudlet ID 2 to VM ID 4 etc.

In this paper, we compared the performance of DEBBP with both conventional binding policy and MaxMin binding policy in each job and total execution time, total execution price, total RAS and QoS. In figure \ref{fig555}, It can be seen that the execution time of the cloudlets in DEBBP is shorter than those of the conventional binding policy and MaxMin based banding policy for both jobs. For job A, the total execution time of DEEBBP was 68.61\% shorter than conventional binding policy and 8.97\% shorter than MaxMin based banding policy. For job B, the total execution time of DEEBBP was 73.84\% shorter than conventional binding policy and 13.53\% shorter than MaxMin based banding policy. In figures 6 and 7, for job A, the total execution price of DEEBBP was 37.16\% less than conventional binding policy and 22.79\% less than MaxMin based banding policy. Similarly, for job B, the total execution price of DEEBBP was 37.64\% less than conventional binding policy, and 23.17\% lower than that of the MaxMin based banding policy. In figures 8 and 9 and for job A, the total RAS of DEEBBP was 47.06\% and 43.31\% higher than conventional binding and MaxMin based banding policies respectvely. For job B, the total RAS of DEEBBP was 44.44\% and 39.87\% higher than the conventional binding and MaxMin based banding policies respectively. In figures 10, for job A, the comprehensive QoS of DEEBBP was significantly higher than the conventional binding policy at 283.12\% and also 65.66\% higher than MaxMin based banding policy. Similarly for job B, the comprehensive QoS was 392.71\% and 61.21\% higher than the conventional binding policy and MaxMin based banding policy respectively.

\begin{table}[H]
\caption{Job A}
\label{tab6}
\begin{tabular}{ccccccccccccc}
\hline
\multirow{2}{*}{\begin{tabular}[c]{@{}c@{}}Cloudlet\\  ID\end{tabular}} & \multicolumn{4}{c}{DEBBP}                                                                                                                                                 & \multicolumn{4}{c}{MaxMin}                                                                                                     & \multicolumn{4}{c}{Conventional}                                                                                                                                          \\
                                                                        & \begin{tabular}[c]{@{}c@{}}VM \\ ID\end{tabular} & \begin{tabular}[c]{@{}c@{}}Exec. \\  Time\end{tabular} & \begin{tabular}[c]{@{}c@{}}Exec.\\   Price\end{tabular} & RAS  & VM ID & \begin{tabular}[c]{@{}c@{}}Exec.\\   Time\end{tabular} & \begin{tabular}[c]{@{}c@{}}Exec.\\   Price\end{tabular} & QoS  & \begin{tabular}[c]{@{}c@{}}VM \\ ID\end{tabular} & \begin{tabular}[c]{@{}c@{}}Exec.\\   Time\end{tabular} & \begin{tabular}[c]{@{}c@{}}Exec.\\   Price\end{tabular} & RAS  \\
                                                                        \hline
0                                                                       & 1                                                & 4.93                                                   & 98.51                                                  & 50   & 1     & 4.93                                                   & 98.51                                                  & 50   & 0                                                & 8.12                                                   & 40.59                                                  & 70   \\
1                                                                       & 2                                                & 3.28                                                   & 32.85                                                  & 40   & 0     & 6.92                                                   & 34.62                                                  & 79   & 1                                                & 1.71                                                   & 34.11                                                  & 50   \\
2                                                                       & 5                                                & 0.47                                                   & 3.28                                                   & 150  & 3     & 3.13                                                   & 15.66                                                  & 110  & 2                                                & 0.8                                                    & 7.98                                                   & 40   \\
3                                                                       & 5                                                & 0.28                                                   & 1.93                                                   & 150  & 2     & 0.47                                                   & 4.69                                                   & 40   & 3                                                & 1.93                                                   & 9.63                                                   & 110  \\
4                                                                       & 5                                                & 1.21                                                   & 8.48                                                   & 150  & 7     & 3.73                                                   & 40.98                                                  & 80   & 4                                                & 7.99                                                   & 39.93                                                  & 60   \\
5                                                                       & 5                                                & 1.71                                                   & 11.94                                                  & 150  & 6     & 8.86                                                   & 53.14                                                  & 70   & 5                                                & 1.71                                                   & 32.11                                                  & 50   \\
6                                                                       & 2                                                & 3.41                                                   & 34.06                                                  & 40   & 5     & 1.95                                                   & 13.68                                                  & 150  & 6                                                & 10.14                                                  & 60.82                                                  & 70   \\
7                                                                       & 1                                                & 1.66                                                   & 33.11                                                  & 50   & 2     & 3.09                                                   & 30.95                                                  & 40   & 7                                                & 5.59                                                   & 61.46                                                  & 80   \\
8                                                                       & 5                                                & 2.26                                                   & 15.81                                                  & 150  & 7     & 6.96                                                   & 76.61                                                  & 80   & 0                                                & 8.12                                                   & 40.59                                                  & 70   \\
9                                                                       & 5                                                & 0.55                                                   & 3.86                                                   & 150  & 6     & 2.89                                                   & 17.32                                                  & 70   & 1                                                & 0.49                                                   & 9.72                                                   & 50   \\
10                                                                      & 0                                                & 10.96                                                  & 54.79                                                  & 70   & 2     & 5.16                                                   & 51.14                                                  & 70   & 2                                                & 5.16                                                   & 51.6                                                   & 40   \\
11                                                                      & 5                                                & 0.83                                                   & 5.78                                                   & 150  & 1     & 0.73                                                   & 14.63                                                  & 50   & 3                                                & 5.55                                                   & 27.75                                                  & 110  \\
12                                                                      & 5                                                & 0.83                                                   & 5.79                                                   & 150  & 5     & 0.83                                                   & 5.78                                                   & 150  & 4                                                & 5.44                                                   & 27.21                                                  & 60   \\
13                                                                      & 5                                                & 0.11                                                   & 0.77                                                   & 150  & 0     & 0.4                                                    & 1.98                                                   & 70   & 5                                                & 0.11                                                   & 0.77                                                   & 150  \\
14                                                                      & 1                                                & 4.27                                                   & 42.66                                                  & 40   & 5     & 4.82                                                   & 33.75                                                  & 150  & 6                                                & 25                                                     & 149.99                                                 & 70   \\
15                                                                      & 5                                                & 1.27                                                   & 8.87                                                   & 150  & 3     & 8.5                                                    & 42.51                                                  & 110  & 7                                                & 3.89                                                   & 42.82                                                  & 80   \\
16                                                                      & 5                                                & 0.91                                                   & 6.36                                                   & 150  & 0     & 3.26                                                   & 16.3                                                   & 70   & 0                                                & 3.27                                                   & 16.37                                                  & 70   \\
17                                                                      & 5                                                & 0.3                                                    & 2.12                                                   & 150  & 4     & 2.06                                                   & 10.32                                                  & 60   & 1                                                & 0.27                                                   & 5.37                                                   & 50   \\
18                                                                      & 3                                                & 10.17                                                  & 50.84                                                  & 110  & 4     & 9.99                                                   & 49.94                                                  & 60   & 2                                                & 2.58                                                   & 25.82                                                  & 40   \\
19                                                                      & 2                                                & 4.27                                                   & 42.66                                                  & 50   & 1     & 2.19                                                   & 43.86                                                  & 50   & 3                                                & 16.66                                                  & 83.32                                                  & 110  \\
\hline
In Total                                                                &                                                  & 11.06                                                  & 507.12                                                 & 2250 &       & 12.15                                                  & 656.88                                                 & 1570 &                                                  & 35.24                                                  & 807.13                                                 & 1530\\
\hline
\end{tabular}
\end{table}

\begin{table}[H]
\caption{Job B}
\label{tab7}
\begin{tabular}{ccccccccccccc}
\hline
\multirow{2}{*}{\begin{tabular}[c]{@{}c@{}}Cloudlet\\  ID\end{tabular}} & \multicolumn{4}{c}{DEBBP}                                                                                                                                                 & \multicolumn{4}{c}{MaxMin}                                                                                                     & \multicolumn{4}{c}{Conventional}                                                                                                                                          \\
                                                                        & \begin{tabular}[c]{@{}c@{}}VM \\ ID\end{tabular} & \begin{tabular}[c]{@{}c@{}}Exec. \\  Time\end{tabular} & \begin{tabular}[c]{@{}c@{}}Exec.\\   Price\end{tabular} & RAS  & VM ID & \begin{tabular}[c]{@{}c@{}}Exec.\\   Time\end{tabular} & \begin{tabular}[c]{@{}c@{}}Exec.\\   Price\end{tabular} & QoS  & \begin{tabular}[c]{@{}c@{}}VM \\ ID\end{tabular} & \begin{tabular}[c]{@{}c@{}}Exec.\\   Time\end{tabular} & \begin{tabular}[c]{@{}c@{}}Exec.\\   Price\end{tabular} & RAS  \\
                                                                        \hline
0                                                                       & 2                                                & 4.37                                                   & 43.65                                                  & 40   & 5     & 2.56                                                   & 17.92                                                  & 150  & 0                                                & 9.2                                                    & 45.99                                                  & 70   \\
1                                                                       & 5                                                & 0.3                                                    & 2.12                                                   & 150  & 1     & 0.27                                                   & 5.37                                                   & 50   & 1                                                & 0.27                                                   & 5.37                                                   & 50   \\
2                                                                       & 5                                                & 0.85                                                   & 5.98                                                   & 150  & 5     & 0.85                                                   & 5.98                                                   & 150  & 2                                                & 1.45                                                   & 14.53                                                  & 40   \\
3                                                                       & 1                                                & 3.17                                                   & 63.39                                                  & 50   & 5     & 3.58                                                   & 25.07                                                  & 150  & 3                                                & 24.07                                                  & 120.35                                                 & 110  \\
4                                                                       & 1                                                & 2.05                                                   & 40.84                                                  & 50   & 7     & 7.11                                                   & 78.25                                                  & 80   & 4                                                & 15.27                                                  & 76.36                                                  & 60   \\
5                                                                       & 5                                                & 0.33                                                   & 2.13                                                   & 150  & 4     & 2.18                                                   & 10.89                                                  & 60   & 5                                                & 0.38                                                   & 2.65                                                   & 150  \\
6                                                                       & 5                                                & 0.85                                                   & 5.98                                                   & 150  & 7     & 2.62                                                   & 28.81                                                  & 80   & 6                                                & 4.43                                                   & 26.56                                                  & 70   \\
7                                                                       & 5                                                & 0.17                                                   & 1.16                                                   & 150  & 7     & 0.51                                                   & 5.59                                                   & 80   & 7                                                & 0.51                                                   & 5.56                                                   & 80   \\
8                                                                       & 1                                                & 5.41                                                   & 108.28                                                 & 50   & 1     & 5.41                                                   & 108.26                                                 & 50   & 0                                                & 21.97                                                  & 109.85                                                 & 70   \\
9                                                                       & 3                                                & 5.36                                                   & 26.82                                                  & 110  & 6     & 4.13                                                   & 24.77                                                  & 70   & 1                                                & 0.71                                                   & 14.11                                                  & 50   \\
10                                                                      & 5                                                & 1.65                                                   & 11.56                                                  & 150  & 6     & 8.57                                                   & 51.39                                                  & 70   & 2                                                & 2.82                                                   & 28.17                                                  & 40   \\
11                                                                      & 5                                                & 2.26                                                   & 15.81                                                  & 150  & 0     & 8.12                                                   & 40.58                                                  & 70   & 3                                                & 15.18                                                  & 75.9                                                   & 110  \\
12                                                                      & 2                                                & 5.4                                                    & 53.95                                                  & 40   & 2     & 5.4                                                    & 53.96                                                  & 40   & 4                                                & 20.9                                                   & 104.52                                                 & 60   \\
13                                                                      & 5                                                & 1.84                                                   & 12.9                                                   & 150  & 2     & 3.14                                                   & 31.43                                                  & 40   & 5                                                & 1.84                                                   & 12.91                                                  & 150  \\
14                                                                      & 7                                                & 7.87                                                   & 86.62                                                  & 80   & 1     & 2.27                                                   & 45.34                                                  & 50   & 6                                                & 31.27                                                  & 14.85                                                  & 70   \\
15                                                                      & 5                                                & 0.72                                                   & 5.01                                                   & 150  & 3     & 4.81                                                   & 24.07                                                  & 110  & 7                                                & 2.2                                                    & 24.2                                                   & 80   \\
16                                                                      & 0                                                & 2.96                                                   & 14.81                                                  & 70   & 0     & 2.97                                                   & 14.83                                                  & 70   & 0                                                & 2.97                                                   & 104.52                                                 & 70   \\
17                                                                      & 5                                                & 1.13                                                   & 7.9                                                    & 150  & 3     & 7.58                                                   & 37.9                                                   & 110  & 1                                                & 1.0                                                    & 19.97                                                  & 50   \\
18                                                                      & 0                                                & 5.02                                                   & 25.11                                                  & 70   & 4     & 9.09                                                   & 45.44                                                  & 60   & 2                                                & 2.34                                                   & 23.45                                                  & 40   \\
19                                                                      & 5                                                & 0.25                                                   & 1.74                                                   & 150  & 2     & 0.42                                                   & 4.23                                                   & 40   & 3                                                & 1.67                                                   & 8.33                                                   & 110  \\
\hline
In Total                                                                &                                                  & 10.73                                                  & 536.05                                                 & 2210 &       & 12.79                                                  & 660.07                                                 & 1580 &                                                  & 41.02                                                  & 813.26                                                 & 1530\\
\hline
\end{tabular}
\end{table}

\subsection*{Implementation of the Proposed VMMNDSA for VM migration}

For the VM migration, we set 4 types of VMs, with each type consisting of 20 different VMs. VMs have the attributes of MIPS, Pes, RAM, Bandwidth, Size. We also set 4 cloud resource providers, each having 10 hosts. The parameters for the hosts are: MIPS, Pes, RAM, Bandwidth and Size as shown on Table \ref{tab888}. This experiment simulates the load balancing experiment of the cloud computing centre through the VMMNDMS dynamic scheduling algorithm of the virtual machine until all tasks are finished. Finally it gives as output - the number of migrations, the average SLA (Service-Level Agreement) violation rate, as well as the overall energy consumption.

\begin{table}[h]
\caption{The experiment parameters of VM migration}
\label{tab888}
\begin{tabular}{cccccccc}
\hline
\multicolumn{2}{c}{VM Setting}                                               & VM Number   & VM MIPS                                               & VM Pes   & VM RAM   & \begin{tabular}[c]{@{}c@{}}VM\\   Bandwidth\end{tabular}   & \begin{tabular}[c]{@{}c@{}}VM\\   Size\end{tabular} \\
\hline
\multirow{4}{*}{\begin{tabular}[c]{@{}c@{}}VM \\ Types\end{tabular}} & Type1 & 20          & 2500                                                  & 1        & 870      & \multirow{4}{*}{100,000}                                   & \multirow{4}{*}{2,500}                              \\
                                                                     & Type2 & 20          & 2000                                                  & 1        & 1740     &                                                            &                                                     \\
                                                                     & Type3 & 20          & 1000                                                  & 1        & 1740     &                                                            &                                                     \\
                                                                     & Type4 & 20          & 500                                                   & 1        & 613      &                                                            &                                                     \\

\hline
\multicolumn{2}{c}{Host Setting}                                             & Host Number & \begin{tabular}[c]{@{}c@{}}Host\\   MIPS\end{tabular} & Host Pes & Host RAM & \begin{tabular}[c]{@{}c@{}}Host\\   Bandwidth\end{tabular} & Host Size \\
\hline
Provider 1                                      & Host Type: 1 & 10          & 1860                                                  & 2        & 4096     & \multirow{2}{*}{100,000,000}                               & \multirow{2}{*}{1,000,000}                          \\
Provider 2 & Host Type: 2& 10          & 2660                                                  & 2        & 4096     &                                                            &                                                     \\
Provider 3 & Host Type: 3 & 10          & 2980                                                  & 2        & 4096     &                                                            &                                                     \\
Provider 4                                                                      & Host Type: 4 & 10          & 3220                                                  & 2        & 4096     &                                                            &                                                     \\

\hline
\multicolumn{2}{c}{the cost of using processing}                             & \multicolumn{6}{c}{3.0}                                                                                                                                                                                      \\
\multicolumn{2}{c}{the cost of using memory}                                 & \multicolumn{6}{c}{0.05}                                                                                                                                                                                     \\
\multicolumn{2}{c}{the cost of using bandwidth}                              & \multicolumn{6}{c}{0.001}                                                                                                                                                                                    \\
\multicolumn{2}{c}{Max/Min Load Threshold}                                   & \multicolumn{6}{c}{80\%/0\%}                                                                                                                                                                                 \\
\hline
\multicolumn{8}{c}{$\alpha = \beta $=0.5}                                                                                                                                                                                                       \\
\hline
\end{tabular}
\end{table}

\begin{figure}[h]
\begin{center}
\begin{subfigure}{0.31\textwidth}
		\includegraphics[width=\textwidth]{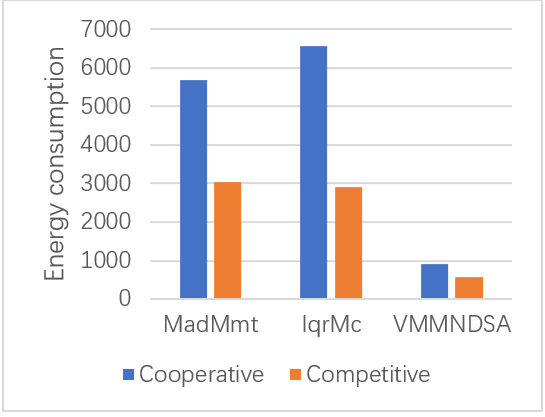}
		\caption{SLA (Service-Level Agreement)}
		\label{fig8a}
	\end{subfigure}
	\begin{subfigure}{0.32\textwidth}
		\includegraphics[width=\textwidth]{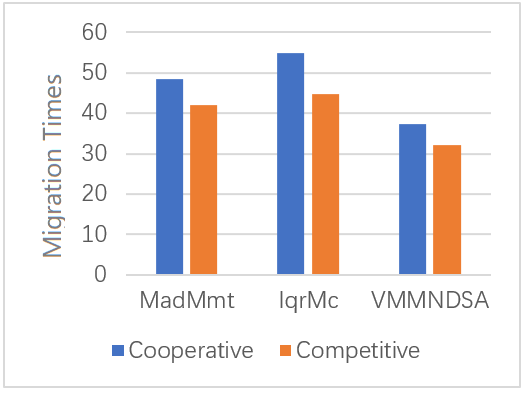}
		\caption{Migration times}	
		\label{fig8b}
	\end{subfigure}
		\begin{subfigure}{0.32\textwidth}
		\includegraphics[width=\textwidth]{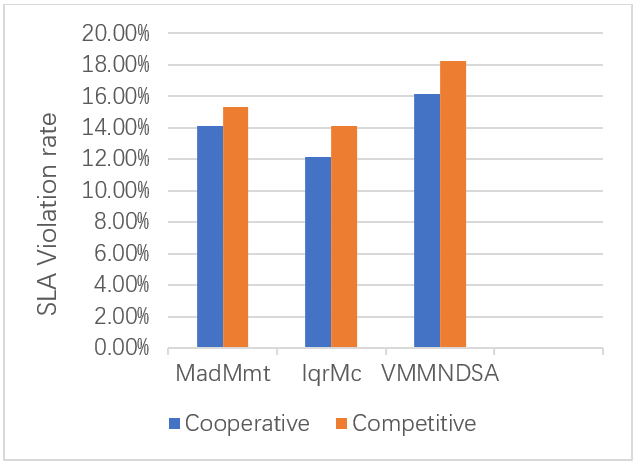}
		\caption{Energy Consumption}
		\label{fig8c}	
	\end{subfigure}
\caption{The execution results of VM migration}
\label{fig8}	
\end{center}
\end{figure}

To verify the performance of VMMNDMS, we compared it with two algorithms: MadMmt and IqrMc \cite{Beloglazov3} and the results are as follows:

Based on the above parameters for evolutionary experiments, we focus on three metrics which are SLA, VM migration times and energy consumption. The results of comparing our proposed VMMNDMS with the MadMmt and IqrMc are shown in Figure \ref{fig8}.

VMMNDMS is aiming at avoiding the unnecessary VM migration, so it has an outstanding advantage of the VM migration times which is showed in figure \ref{fig8a} in both cooperative and competitive federation where it is 84.05\% and 80.89\% less than MadMmt and where it is 86.20\% and 80.02\% less than IqrMc. 

At the same time, figure \ref{fig8b} shows that VMMNDMS is the most efficient one since it has the lowest energy consumption in both cooperative and competitive federation where it resulted is 22.99\% and 32.18\% lower energy compared to MadMmt, and 23.72\% and 28.52\% lower compared to IqrMc. 

In our model, SLA relates to the number of hot nodes, since our strategy focusing on avoiding the unnecessary VM migrations to save energy; it has a relative higher SLA violation rate compared to MadMmt and IqrMc and this is as shown in figure \ref{fig8c}, for cooperative and competitive federation, which are respectively 14.60\%, 33.41\% and 19.16\%, 29.31\%. 

We also compared the cooperative federation and competitive federation in Table \ref{tab9}. For migration times, the competitive federation is better, because it has less choice of target nodes. For energy consumption however, cooperative federation plays better its VMs have more selections options but the migrations are more extensive. Comparatively, in our model the performance of SLA and energy are opposite,thus competitive federation results in better SLA adherance.

\begin{table}[h]
\caption{The Comparison of Cooperative Federation and Competitive Federation}
\label{tab9}
\resizebox{\textwidth}{!}{%
\begin{tabular}{|l|llllll|}
\hline
 & \multicolumn{3}{l|}{Cooperative Federation} & \multicolumn{3}{l|}{Competitive Federation} \\ \hline
 & MadMmt & IqrMc & VMMNDSA & MadMmt & IqrMc & VMMNDSA \\ \hline
Migration times (less) & \multicolumn{3}{l|}{} & 46.52\% & 55.73\% & 35.91\% \\ \hline
Energy Consumption (less) & 13.28\% & 18.50\% & 14.11\% & \multicolumn{3}{l|}{} \\ \hline
SLA (less) & \multicolumn{3}{l|}{} & 7.72\% & 13.98\% & 12.68\% \\ \hline
\end{tabular}%
}
\end{table}

Based on the above analysis, it can be concluded that our proposed VMMNDMS can greatly reduce the number of virtual machine migrations, so as to reduce the impact of migration on the overall performance of the cloud computing centre and achieve more energy-efficient results, while keeping SLA in an acceptable level. The above results demonstrate the efficient of our proposed VMMNDMS.

\section*{Conclusion}
Under the premise of the same hardware facilities, the level of service quality in cloud computing largely depends on the resource allocation strategy adopted. In this paper, based on the structure of different levels of resources in the cloud computing environment, we proposed a two-in-one solution model; at first, we assign user workloads to the virtual machine for the task assignment between the application layer and the virtual machine layer. We then use the DE algorithm to implement the proposed multi-based QoS binding strategy. We verified the efficiency of through results obtained from simulations conducted using the CloudSim toolkit. The second part of the model, focuses on the virtual machine and physical layers, with the aim of striking a balance between energy consumption and SLA. We proposed an effective scheduling algorithm that takes virtual machine migration cost and host load assessment into consideration when migrating workloads. 
In the future, this work can be extended to consider factors external to the Cloud, particularly the networks \cite{Zennaro}. This is imperative because the network can severely impact access to cloud nodes as well as affect the QoS provided by the cloud or perceived by the user. Furthermore, obtained results were based on simulations carried out using CloudSim, implementing these on real testbeds such as OpenStack could be another potential area for future research work.


\end{document}